%% file: main.tex
\documentclass[a4paper,11pt]{article}
\pdfoutput=1

\usepackage{jinstpub} 
\usepackage[utf8]{inputenc}
\usepackage[english]{babel}
\usepackage[ruled,vlined,linesnumbered]{algorithm2e} 
\usepackage[noend]{algpseudocode}
\usepackage{float}
\usepackage[justification=justified,labelfont=bf]{caption}
\usepackage[nottoc,numbib]{tocbibind}  
\usepackage{tikz}
\usetikzlibrary{fit,calc}
\usepackage{balance} 
\usepackage{mathtools}
\usepackage{amsmath}
\usepackage{subcaption}
\usepackage{pgfplots}
\usepackage{booktabs}
\usepackage{siunitx}
\usepackage{natbib}
\usepackage{graphicx}

\colorlet{myred}{red!85!black}
\colorlet{myblue}{blue!80!black}
\colorlet{mycyan}{cyan!80!black}
\colorlet{mygreen}{green!70!black}
\colorlet{myorange}{orange!90!black!80}
\colorlet{mypurple}{red!50!blue!90!black!80}
\colorlet{mydarkred}{myred!80!black}
\colorlet{mydarkblue}{myblue!80!black}
\colorlet{mypink}{red!40}
\colorlet{mylightcyan}{cyan!60}

\DeclareSIUnit{\dBm}{dBm}
\DeclareSIUnit{\bit}{bit}
\DeclareSIUnit{\byte}{B}
\DeclareSIUnit{\eV}{eV}

\DeclareMathAlphabet{\mathpzc}{OT1}{pzc}{m}{it}

\SetCommentSty{mycomment}
\DontPrintSemicolon
\SetAlgoLined
\DontPrintSemicolon

\newcommand{\boxit}[2]{
	\tikz[remember picture,overlay] \node (A) {};\ignorespaces
	\tikz[remember picture,overlay]{\node[yshift=3pt,fill=#1,opacity=.25,fit={($(A)+(0,0.15\baselineskip)$)($(A)+(0.945\linewidth,-{#2}\baselineskip - 0.25\baselineskip)$)}] {};}\ignorespaces
}

\newcommand{\uMUX}{$\mu$MUX}

\graphicspath{{figuras/}{figuras/mmb/}{figuras/logos/}{figuras/backend_diagrams/}{figuras/setup_med/}{figuras/simus_python_ddc/}{figuras/simus_vivado_ddc/}{figuras/ddc/}{figuras/signal_flow/}{figuras/fpga/}{figuras/goertzel/}{figuras/sensors/}{figuras/measurements/}{figuras/generation/}}

\title{An Implementation of a Channelizer based on a Goertzel Filter Bank for the Read-Out of Cryogenic Sensors}

\author[a,b,c,d]{L. P. Ferreyro}
\author[a,b,c,d]{M. García Redondo}
\author[a,d,f]{M. R. Hampel}
\author[a,d,f]{A. Almela}
\author[a,d,f]{A. Fuster}
\author[a,b,c,d]{J. Salum}
\author[a,b,e,d]{J. M. Geria}
\author[f]{J. Bonaparte}
\author[a,b]{J. Bonilla-Neira}
\author[a,b]{N. Müller}
\author[b,c]{N. Karcher}
\author[b,c]{O. Sander}
\author[a]{M. Platino}
\author[b,c]{M. Weber}
\author[a]{A. Etchegoyen}

\affiliation[a]{Instituto de Tecnologías en Detección y Astropartículas (CNEA - CONICET - UNSAM), Buenos Aires, Argentina}
\affiliation[b]{Karlsruhe Institute of Technology (KIT), Karlsruhe, Germany}
\affiliation[c]{Institute for Data Processing and Electronics (IPE), Karlsruhe, Germany}
\affiliation[d]{Universidad Tecnológica Nacional - Facultad Regional Buenos Aires, Argentina}
\affiliation[e]{Institute for Micro- and Nanoelectronic Systems (IMS), Karlsruhe, Germany}
\affiliation[f]{Comisión Nacional de Energía Atómica (CNEA)}

\emailAdd{luciano.ferreyro@iteda.cnea.gov.ar}

\keywords{Instruments for CMB observations, Spectrometers, Digital electronic circuits, Digital signal processing (DSP)}

\input{./sections/abstract.tex}

\begin{document}

	\maketitle
	
	\flushbottom
 
	\input{./sections/introduction.tex}

\input{./sections/readout_concept.tex}
%
    \input{./sections/goertzel_bank_theory.tex}
	\input{./sections/validation.tex}
	\input{./sections/conclusions.tex}
		
	\bibliographystyle{JHEP}
	\bibliography{main}

\end{document}

%% file: sections/abstract.tex
\abstract{
In this work we present an application of the Goertzel Filter for the channelization of multi-tonal signals, typically used for the read-out of cryogenic sensors which are multiplexed in the frequency domain (FDM), by means of Microwave Superconducting Quantum Interference Device (SQUID) Multiplexer (\uMUX). We demonstrate how implementing a bank of many of these filters, can be used to perform a channelization of the multi-tonal input signal to retrieve the data added by the sensors. We show how this approach can be implemented in a resource-efficient manner in a Field Programmable Gate Array (FPGA) within the state-of-the-art, which allows great scalability for reading thousands of sensors; as is required by Radio Telescopes in Cosmic Microwave Background Radiation (CMB) surveys using cryogenic bolometers, particles detection like Neutrino mass estimation using cryogenic calorimeters or Quantum Computing.
}

%% file: sections/introduction.tex
\section{Introduction}
\label{sec:intro}
There are many unanswered questions in the scientific community. In particle physics, research into the absolute scale of neutrino masses is of great interest and has yet to be determined. Recently, the KATRIN project \cite{Aker2022,Weinheimer2002} established a new upper limit of \SI{0.8}{eV /c^{2}} by studying the beta decay of tritium and measuring the energy of the electrons released in the process. Using a different approach, the ECHo project \cite{Gastaldo2017} aims to achieve sub-eV sensitivity for the effective electron neutrino mass by analysing the calorimetric electron capture spectrum of \textsuperscript{163}Ho.
Another example of still unanswered questions are those related to the Cosmic Microwave Background Radiation (CMB). The CMB was predicted in 1948 by Ralph Alpher and Robert Herman, and first measured in 1964 by Arno Penzias and Robert Woodrow Wilson using a radiometer originally designed for radio astronomy and satellite communications. They measured an unexpected spectra that matched a black body radiating at \SI{3.5}{K} \cite{Penzias1965}, which was indeed accounted for by the CMB. More recently, results from the Cosmic Background Explorer (COBE) \cite{Mather1991} showed that \SI{50}{\%} of the luminosity and \SI{98}{\%} of the photons emitted by the CMB are in the sub-millimetre and far-infrared range. Experiments so far suggest a big bang model for a hot and dense universe in the past, cooling adiabatically as it expands, and the CMB represents the oldest picture, the last scattering, we can take of the universe when it was only $\sim$380,000 years old. A period of exponential expansion called "inflation" \cite{Spergel1997,Seljak1997} in the early universe has been proposed as a solution to major problems with the standard big bang model. One observable effect of this process is the production of B-modes in the CMB. Experiments such as LABOCA \cite{Siringo2009}, Bicep2 \cite{Ade2014}, Boomerang and Maxima \cite{Bond2000}, SPT \cite{Carlstrom2011}, and now QUBIC\cite{Hamilton2022} aim to measure the B-modes.
Both the neutrino mass experiments and the CMB measurements implement low-temperature detectors: calorimeters and bolometers, respectively. These detectors operate at very low temperatures, typically below \SI{1}{K}. For neutrino mass experiments, the detectors must guarantee high energy resolution, fast response and a quantum efficiency close to 100 \%; for ECHo, this is achieved by using Metallic Magnetic Calorimeters \cite{Fleischmann2005,Fleischmann2009} (MMC). In CMB experiments, it is desirable to work with background-limited detectors, such as the most commonly used Transition Edge Sensors (TES) \cite{Irwin2005}. 

The direct connection of each sensor to a read-out electronics is possible as long as the number of sensors is kept in a low value, mainly due to the complexity of having many cables coming out of the cryostat. The need to improve the sensitivity of the experiment requires an increase in the number of sensors, making this direct connection no longer suitable as the complexity added by the large number of cables required increases considerably. Two different techniques are used to solve this problem: Time Division Multiplexing (TDM), used i.e. in QUBIC and Frequency Division Multiplexing (FDM), used i.e. in ECHo. Both techniques reduce the number of cables needed to read a large number of sensors.

The growing demand for larger scales detector arrays in these experiments drives advances not only in the low temperature devices but also in the associated room temperature electronics. The latter involves the challenging tasks of generating the sensors monitoring signals, their acquisition and processing. In light of this, we present a method based on a Goertzel Filter Bank for the read-out electronics of frequency division multiplexed cryogenic detectors. This method is suitable for any of the aforementioned experiments due to its inherent versatility in coarse and fine tuning of the channelization process, and its signal demodulation capabilities are evaluated.


%% file: sections/readout_concept.tex
\section{The read-out system}
\label{sec:readconcept}

\begin{figure}[H]
	\centering
	\includegraphics[width=0.85\textwidth]{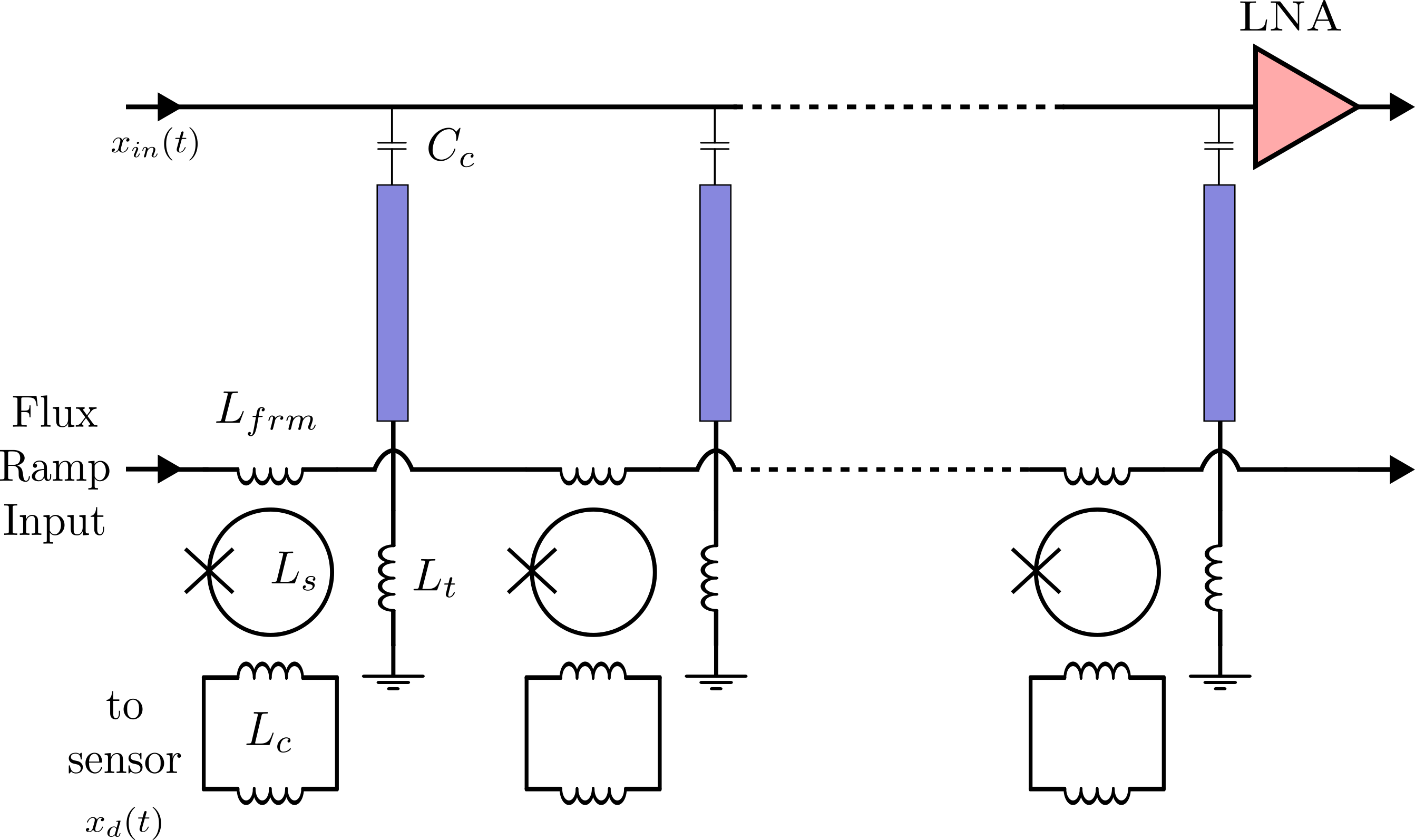}
	\caption{Microwave Superconducting Quantum Interference Device Multiplexer schematic with the FRM line. This device is placed inside of a cryostat typcally between \SI{10}{m\kelvin} and \SI{100}{m\kelvin}. $C_c$ is the coupling capacitor, $L_t$ and the blue rectangle represents the superconducting microwave resonator, $L_s$ represents the rf-SQUID inductance, $L_{frm}$ the coupling inductor between the FRM lane and the rf-SQUID, $L_c$ is the coupling inductor to the required sensor and LNA stands for Low-Noise Amplifier.}
	\label{fig:umux_flux_ramp}
\end{figure}

The Microwave Superconducting Quantum Interference Device (SQUID) Multiplexer (\uMUX{}) enables the read-out of large arrays of cryogenic sensors \cite{Irwin2004, Jaklevic1964, Hirayama2013}, combining the signals from those sensors onto a pair of coaxial cables. Each input channel of the \uMUX{} consists of a non-dissipative dc- or rf-SQUID coupled to a superconducting microwave resonator with a unique resonance frequency, and each channel is capacitively coupled to a common microwave feed line, see figure \ref{fig:umux_flux_ramp}. The magnetic flux within the SQUID modifies the resonance frequency of the associated resonantor by changing its inductance, due to a variation in its coupled sensor. By reading this frequency shift one can measure the variation of magnetic flux in the SQUID and therefore the sensor signal. In this work, we focused on a system with an rf-SQUID multiplexer for a Frequency Division Multiplexing scheme.

The linear read-out of the \uMUX{} is carried out by implementing the Flux-Ramp Modulation (FRM) technique, which consists of applying a periodic ramp signal to all the SQUIDs. The FRM may be considered as a phase modulation technique because the SQUIDs have a periodic response, in which case the input signal determines the instantaneous phase of the periodic response to the flux ramp. Even, if the SQUIDs response is essentially sinusoidal it may also by considered a frequency modulation technique \cite{Mates2012}.

Against this, a practical way of measuring the resonator's frequency is to inject a carrier signal, $x_{in}(t)$, into the multiplexer line and analyse how it changes, because the signals information therefore appears in the sidebands of $x_{in}(t)$.

\begin{figure}[H]
	\centering
	\includegraphics[width=0.75\textwidth]{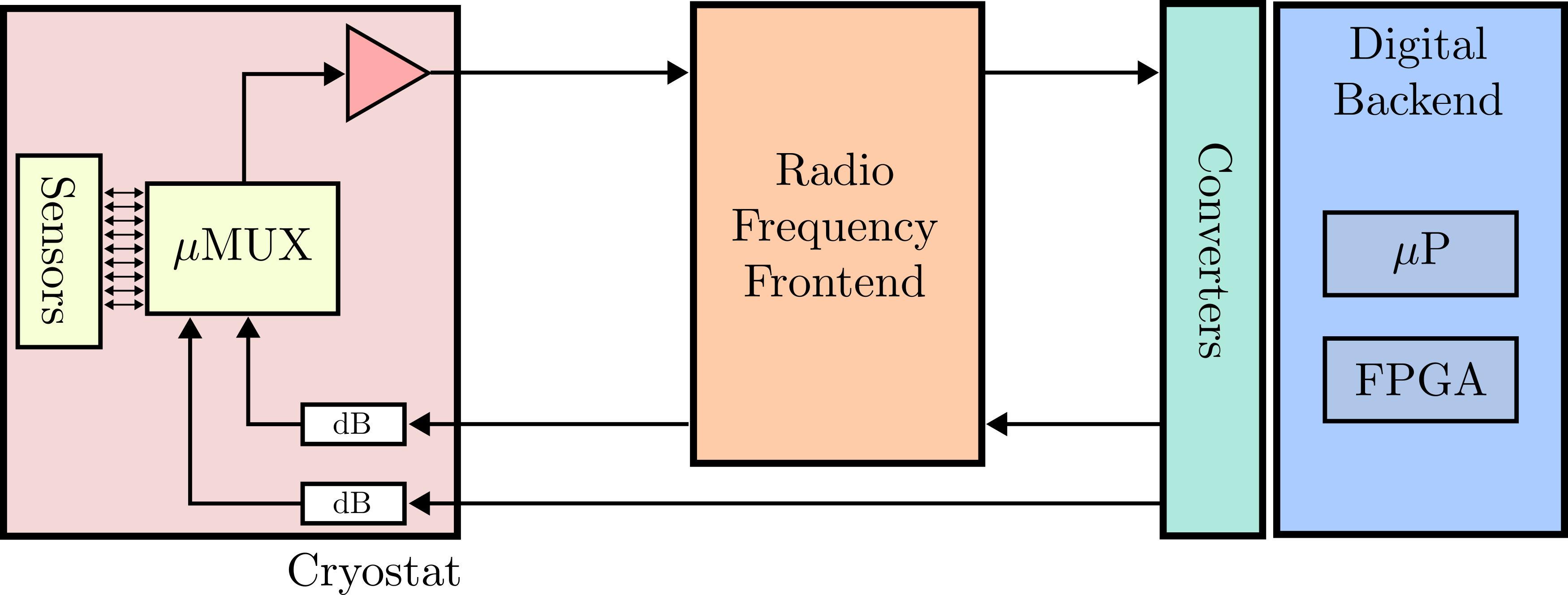}
	\caption{The read-out system. The cryostat has of different temperature stages, not sketched in this diagram. The LNA, represented by the red triangle, works in the \SI{4}{\kelvin} stage. The \uMUX~and the sensors are working in the lowest temperature stage. }
	\label{fig:read_out_system}
\end{figure}

The read-out system presented in Figure \ref{fig:read_out_system} achieves its functionality through two sets of electronics: the cold electronics and the room temperature electronics. The cold electronics, already introduced, mainly consists of sensors and the \uMUX{}. On the other hand, the room temperature electronics include a Digital Backend, which houses a Field Programmable Gate Array (FPGA) and a micro-processor ($\mu$P). The FPGA implements the communication protocols with the high-performance and high sampling rate converters, Analog-to-Digital Converter (ADC) and Digital-to-Analog Converter (DAC), performs the IQ modulated generation of $x_{in}(t)$ in base-band (BB) and the FRM signal, the acquisition of the resultant signal after the cryostat and the Radio-Frequency Frontend, and the pre-processing of this signal. The $\mu$P is responsible for system control, final processing and data delivery. The Radio Frequency Frontend, consists of a transmission path for the up-conversion of $x_{in}(t)$ to the \uMUX~operating region (between \SI{4}{\giga\hertz} and \SI{8}{\giga\hertz}), and a reception path for the down-conversion of $x_{in}(t)$ to BB again in order to be read by the ADCs in the converters block.

The main focus of this study is in the development of the Digital Backend. The used hardware consists of a Xilinx's ZCU102 and an Analog Devices AD-FMCDAQ2-EBZ board composed by an AD9680 ADC and an AD9144 DAC, both sampling at \SI{1}{Gsps}. The ADC has been configured to use four Digital Down Converters (DDC), giving a sampling frequency of \SI{250}{Msps} in the FPGA side.

We propose the demodulation of the input signal, $x_{in}(t)$, by calculating the desired Discrete Fourier Transform (DFT) bin consistent with each monitoring signal, while in a further step demodulate the sensor signal component, $x_{d}(t)$. One way we can efficiently compute a single DFT bin is using the Goertzel Filter.

%% file: sections/goertzel_bank_theory.tex
\section{The Goertzel Filter Bank Channelizer}
\label{sec:goertzeltheory}

\subsection{The Goertzel Filter}

The Goertzel Algorithm (GF) was introduced in 1950 by Gerald Goertzel \cite{Goertzel1958} 
to calculate a single bin of the Discrete Fourier Transform (DFT), the \textit{kth} bin of an N-point DFT, as defined in \cite{Sysel2012}:

\begin{equation}
    X[k] = \sum_{n=0}^{N-1}x[n]e^{-j2\pi k\frac{n - N}{N}} 
    \label{eq:gf_definition}
\end{equation}

which can be implemented as a second order Infinite Impulse Response (IIR) filter.



Looking at \eqref{eq:gf_definition} we see that $x[n]$ is multiplied by a rectangular window $w[n]$. Generalizing this expression we get: 

\begin{equation}
    X[k] = \sum_{n=0}^{N-1}w[n]x[n]e^{-j2\pi k\frac{n - N}{N}}
    \label{eq:fourier_disc_euler_win}
\end{equation}

\begin{figure}[H]
    \centering
    \includegraphics[width=1\columnwidth]{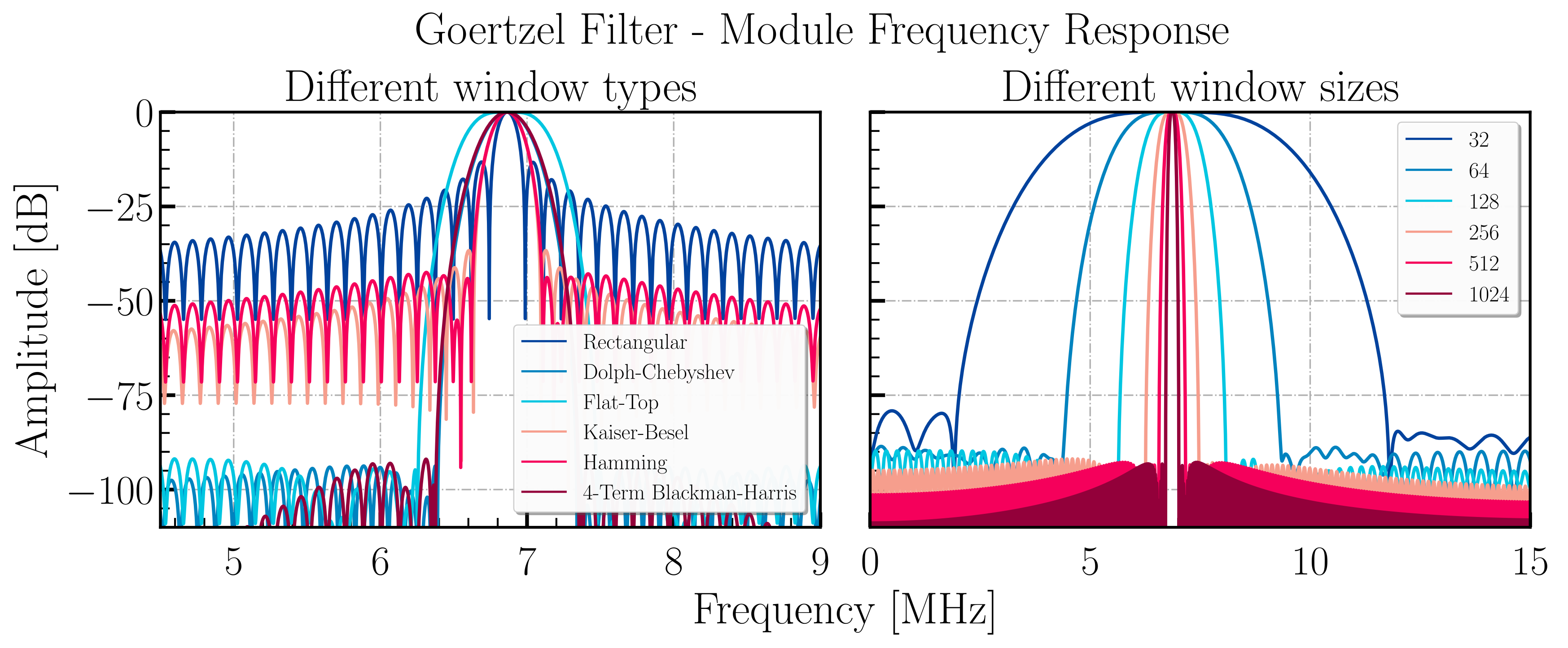}
    \caption{\textbf{Left:} the Goertzel Filter response for different window types with a fixed size of 256 samples. \textbf{Right:} the response for a Flat-Top window for different sizes. The other simulation parameters were: a sampling frequency of \SI{31.25}{MSPS} and the GF centered in \SI{6.865}{MHz}.}
    \label{fig:gf_response_diff_windows}
\end{figure}

In the frequency domain, we have the convolution of $W[k]$ (which is $\mathcal{F}\{w[n]\}$) with the GF spectral response, which is essentially a Kronecker delta. Windowing functions help in this application in three important aspects: a) mitigating the spectral leakage effects due to the inherent windowing procedure of applying a Fourier transform method, b) improving the isolation between adjacent channels, and c) filtering spurious and intermodulation products. The designed firmware allows any desired windowing function to be configured by software. A modified version of the Flat-Top window is used as the main window (but others can be used, \cite{Harris1978,Heinzel2002}) in order to achieve: an acceptable flatness in the band-pass, a strong attenuation starting from the Highest Side Lobe (HSL) of at least \SI{-100}{\decibel} and high enough processing gain (PG). Figure \ref{fig:gf_response_diff_windows} shows some windows convolved with the Goertzel Filter. 

\subsection{Firmware implementation}

In order to implement a bank of digital filters with as many as possible to channelize multiple input tones, the GF must be efficiently mapped to the FPGA of the Digital Backend (DB), as shown in figure \ref{fig:read_out_system}. We looked for an implementation that would reduce the number of computations, allow complex multiplications and be able to compute non-integer multiples of the fundamental frequency. The solution proposed by \cite{Sysel2012} meets these requirements and is presented in the algorithm \ref{alg:goertzel_2}.

\begin{algorithm}[H]    
    \caption{Goertzel algorithm generalized to non-integer multiples of fundamental frequency}
    \label{alg:goertzel_2}
    $\alpha = \frac{2\pi k}{N}$\;
    $\beta = \frac{2\pi k(N-1)}{N}$\;
    $a = \cos(\beta)$\;
    $b = -\sin(\beta)$\;
    $c = \sin(\alpha)\sin(\beta) - \cos(\alpha)\cos(\beta)$\;
    $d = \sin(2\pi k)$\;
    $\omega_0 = \omega_1 = \omega_2 = 0$\;
        \boxit{mylightcyan}{3.2}
        \For(\Comment{Iterative Section}){$window_{index}=0$ to $window_{size}-1$}{ 
            $\omega_0 = x_{in} + 2 \cos(\alpha) \omega_1 - \omega_2$\; 
            $\omega_2 = \omega_1$\;
            $\omega_1 = \omega_0$\;
            }
    \boxit{mypink}{2.35}
    $X_k = a\omega_1 + c\omega_2 + j(b\omega_1 + d\omega_2)$ \hspace{5cm} \Comment{Non-iterative Section}\; 
    $|X_k| = \sqrt{(a\omega_1 + c\omega_2)^2 + (b\omega_1 + d\omega_2)^2}$\;
    $\Phi(X_k) = \arctan(\frac{b\omega_1 + d\omega_2}{a\omega_1 + c\omega_2})$\;
\end{algorithm} 
\vspace{0.25cm}

\begin{figure}[H]
	\centering
        \includegraphics[width=0.7\columnwidth]{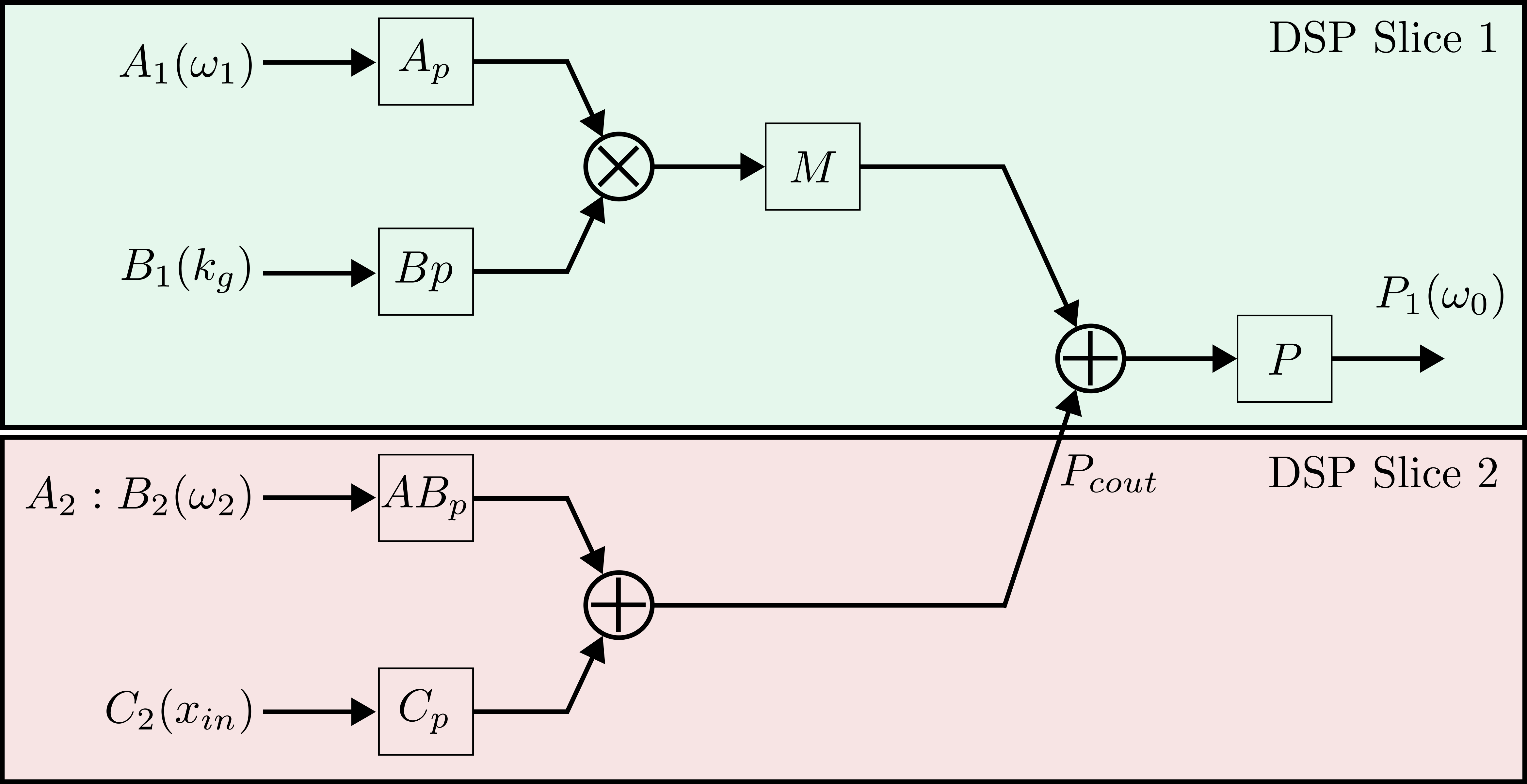}		      
        \caption{Goertzel Filter DSP Slices mapping: A\textsubscript{p}, B\textsubscript{p}, AB\textsubscript{p}, C\textsubscript{p}, M and P refers to the pipeline level within the DSP Slice configuration. As $P_{cout}$ is configured to be used, the two DSP Slices are neighbors in the chip. This is applied to all the used DSP Slices involved in the GF mapping.}
        \label{fig:gf_dsp_slices_implementation}
\end{figure}

We split the algorithm into two main parts, an iterative section and a non-iterative section. From the iterative part, $\omega_0 = x_{in} + 2 \cos(\alpha) \omega_1 - \omega_2$, was mapped to 2 DSP Slices and their interconnection can be appreciated in figure \ref{fig:gf_dsp_slices_implementation}. Two clock cycles are required for determining one $\omega_0$ value which in principle states that the ratio between the logic operation frequency, $f_{logic}$, and the sampling frequency of the incoming samples, $F_S$, needs to be at least two.

However, adding the appropriate pipeline stages to the architecture removes this limitation, allowing $F_S = f_{logic}$. 

The chosen ADC for the prototype operates at \SI{1}{GSPS} (IQ sampling). For this implementation we decided to work with $f_{logic}$ = \SI{250}{\mega\Hz}. In addition, to save resources, we also used one of the ADC features that implements a decimation by 4, giving a resulting $F_s$ of \SI{250}{MSPS}. 

\begin{figure}[H]
	\centering
		\includegraphics[width=0.85\columnwidth]{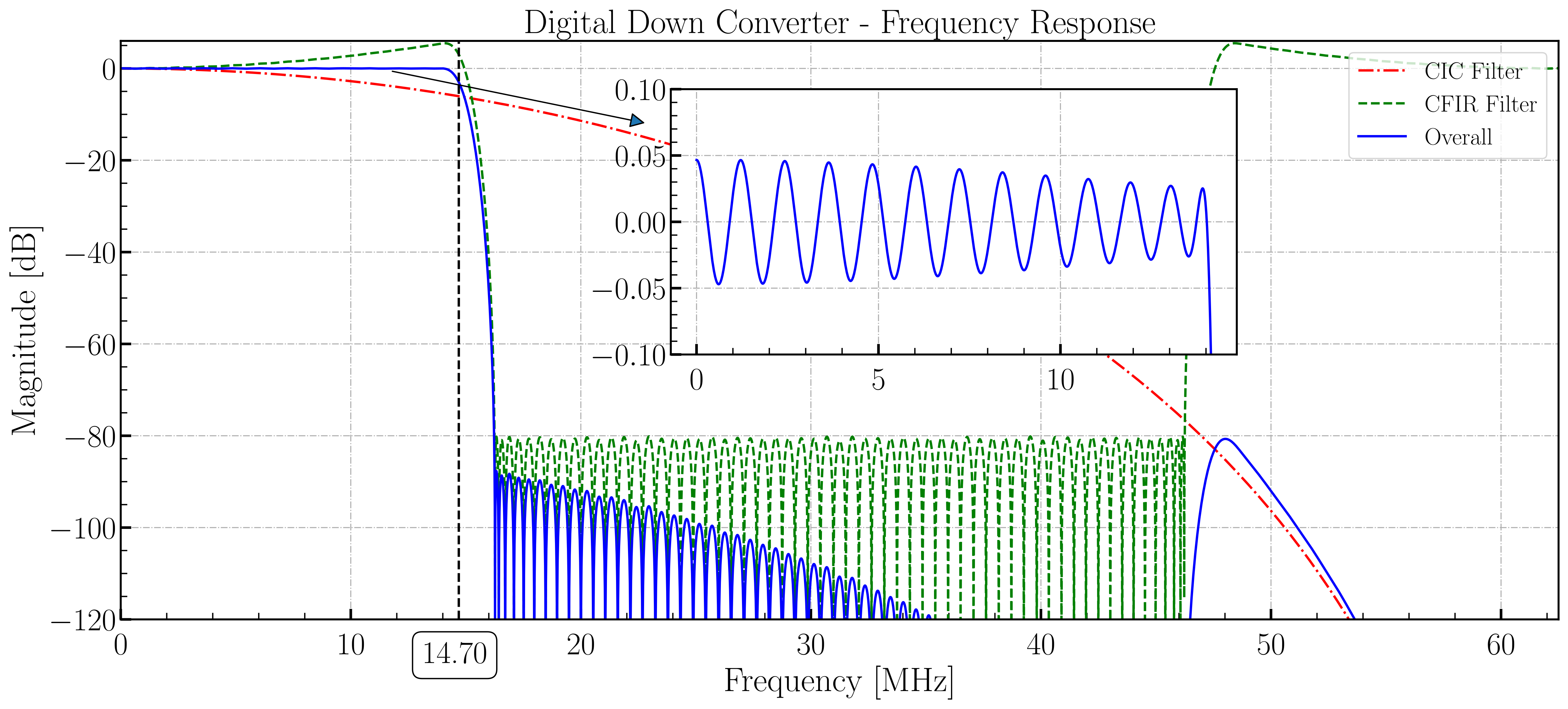}		
        \caption{Digital Down Converter: simulated frequency response of the whole DDC chain where the CIC, CFIR and overall response are plotted. The \SI{3}{\dB} frequency cut, $f_{cut}$, marked with the dashed black vertical line is \SI{14.70}{\mega\Hz} and a ripple of less than 0.1 dB can be observed.} 
        \label{fig:ddc_chain_model}
\end{figure}

\begin{figure}[H]
	\centering
	\includegraphics[width=0.8\columnwidth]{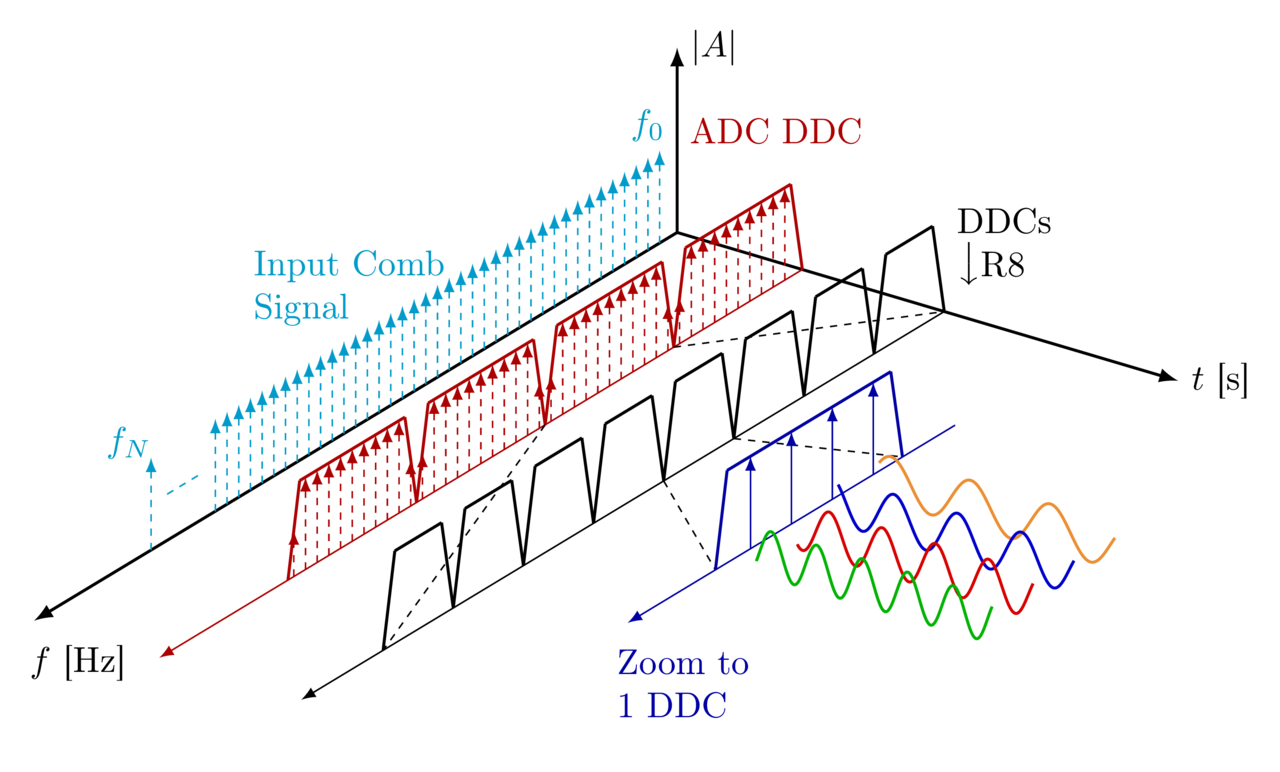}
	\caption{Channelization process from the ADC to the Digital Down Converter stages in the proposed firmware.}
	\label{fig:acq_chain_process}
\end{figure}

Finally, we implemented a second decimation stage by 8 using a Digital Down Converter (DDC), resulting in a final sampling frequency of \SI{31.25}{MSPS}. The implemented DDC is based on a Cascaded Integrator-Comb (CIC) filter \cite{Hogenauer1981} and its simulated frequency response is shown in figure \ref{fig:ddc_chain_model}. It uses 3 DSP slices for the complex mixer and 16 DSP slices for the Compensation FIR (CFIR) filter (considering the two channels, I and Q). The decision to use a second DDC stage is based on exploiting and exploring the combination of the defined $f_{logic}$ and the need to introduce a register in the feedback path of the Goertzel Filter. The channelization process from the ADC to the DDC outputs described so far, can be seen in figure \ref{fig:acq_chain_process}. The output is fed into the Goertzel Filter Bank (GFB) for the final channelization step. This setup is capable of operating within a complex bandwidth of \SI{720}{MHz} \cite{analog:ad9680}.

The GF structure presented in figure \ref{fig:gf_dsp_slices_implementation} only works for one signal component when $F_S = f_{logic}$ (no second DDC stage). When working with complex signals in quadrature modulation, two structures are required (one for I and one for Q), a total of four DSP slices:

\begin{equation}
    GF_{DSP_{0}} = 4N_{CT}
    \label{eq:gf_no_ddc_dsps_slices_use}
\end{equation}

where N\textsubscript{CT} is the number of Complex Tones to process. When the GF is combined with the DDC stage, the expression to calculate the number of DSP Slices used is: 

\begin{equation}
    GF_{DSP_{1}} = \frac{4N_{CT}}{R} + R(\frac{128}{R} + 3) = \frac{4N_{CT}}{R} + 3R + 128
    \label{eq:gf_ddc_dsps_slices_use}
\end{equation}

where R stands for \textit{Decimation Ratio}. $R(\frac{128}{R} + 3)$ represents the relationship between the required DSP Slices in the Compensation FIR, the input $F_s$ (after the CIC filter), $f_{logic}$ = \SI{250}{\mega\hertz}, the filters order and the quantization parameters, according to Xilinx's FIR Compiler tool \cite{fir_compiler}. The optimum value for this expression is determined by analysing its derivative:

\begin{equation}
    \frac{\partial GF_{DSP_{1}}}{\partial R} = -\frac{4N_{CT}}{R^2} + 3
    \label{eq:gf_ddc_dsps_slices_use_optim}
\end{equation}

\begin{figure}[H]
	\centering
	\includegraphics[width=0.75\columnwidth]{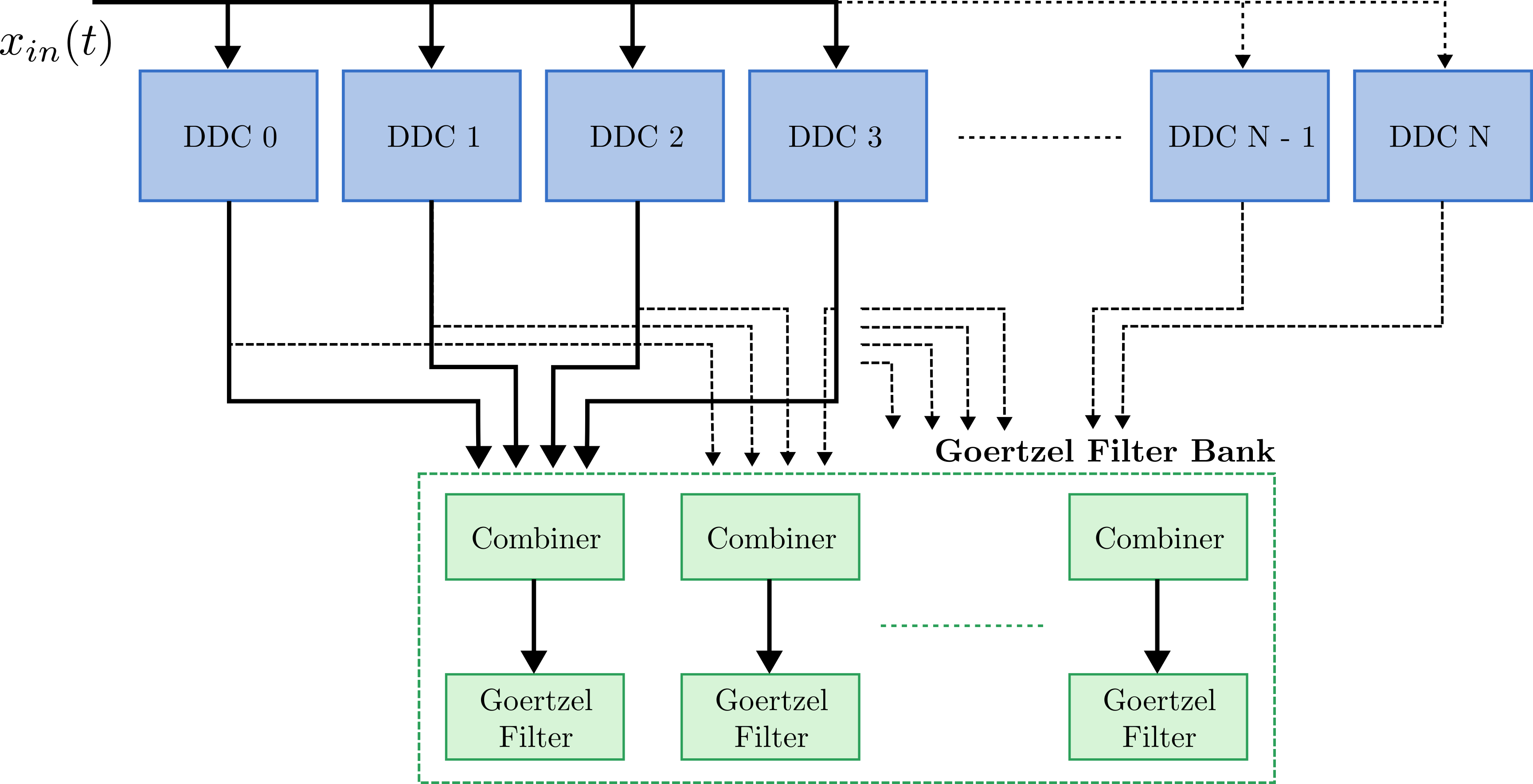}
	\caption{Block Diagram of the interconnection between DDCs and GF Cores (the core is composed by the combiner and the GF). Each GF core handles 4 DDCs. The combiner will collect the samples for the DDCs and serialize them to the GF. Each DDC is connected to several GF cores.}
	\label{fig:ddc_to_gf}
\end{figure}

For example, for a fixed value of N\textsubscript{CT} = 64, GF\textsubscript{DSP\textsubscript{0}} = 256 DSP slices, however the optimum R value in \eqref{eq:gf_ddc_dsps_slices_use_optim} is $\sim9.24$, resulting in 183.4 DSP Slices. R = 8 is the closest feasible value to work with \cite{Hogenauer1981}. Adding this decimation ratio introduces the capability of the designed GF architecture of figure \ref{fig:gf_dsp_slices_implementation} to process 4 complex DFT bins (or 8 real DFT bins). This is achieved by serially supplying the samples from 4 different DDCs. The structure is depicted in figure \ref{fig:ddc_to_gf}.


\begin{figure}[H]
	\centering
	\includegraphics[width=0.85\columnwidth]{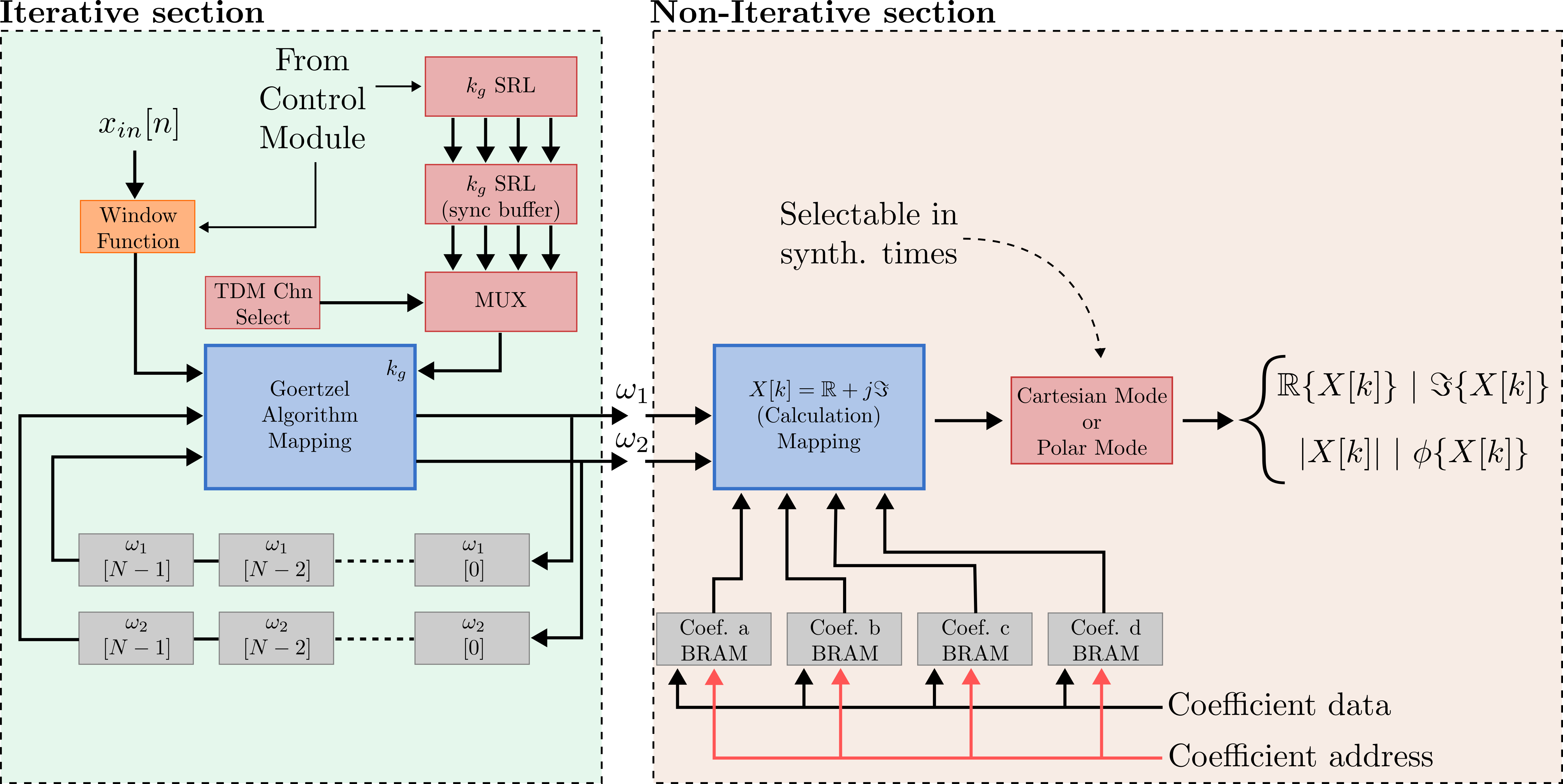}
	\caption{Block Diagram of the implemented Goertzel Filter structure.}
	\label{fig:gf_mapping}
\end{figure}

The GF block in figure \ref{fig:ddc_to_gf} is displayed in detail in figure \ref{fig:gf_mapping}. The Window Function block cyclically applies the desired window to the input signal, $x_{in}$, and it is a common block for all the GF cores. The module that calculates $X[k]$ in the Non-Iterative section of the figure \ref{fig:gf_mapping} uses 12 DSP Slices and handles 32 IQ components when connected to eight Goertzel Algorithm Mapping modules. 

\subsection{Window function compensation}
\begin{figure}[H]
	\centering
	\includegraphics[width=0.7\textwidth]{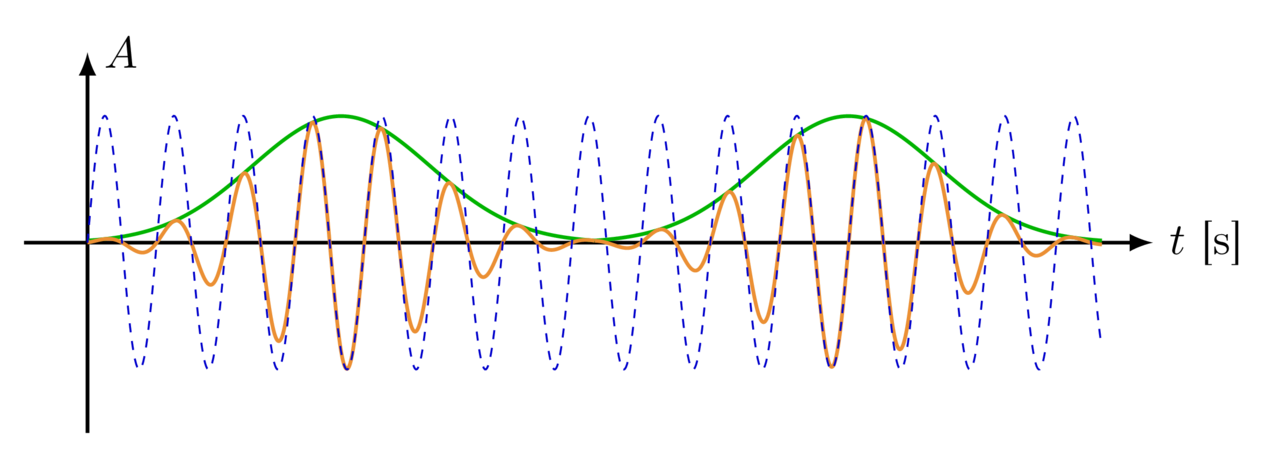}
	\captionof{figure}{Window function (green) applied to a signal (dashed blue).}
	\label{fig:win_function_pipeline}
\end{figure}

The use of window functions to improve the channelizer performance also distorts the signal as can be seen in figure \ref{fig:win_function_pipeline}, where it is possible to appreciate that the energy and the amplitude of the signal were affected. Since in this work the scientific data is retrieved by analyzing the amplitude of the input tones the amplitude compensation is necessary. 

This can be corrected by performing the following calculation, which makes the process compatible with any type of window: 

\begin{equation}
    ACF = \frac{1}{\sum_{i = 0}^{N - 1} w[n]}
\end{equation}

where $ACF$ is the window \textit{Amplitude Correction Factor}, which is also the inverse of the \textit{Coherent Gain}. This compensation is applied in the non-iterative section after the real and imaginary parts of $X[k]$ are calculated.

\subsection{Arithmetic treatment}
One of the major considerations during the algorithm implementation is the fixed-point arithmetic treatment. As the GF is an IIR Filter type, the arithmetic growth within the internal registers must be controlled. Several works \cite{Beraldin1989,medina2009,medina2012} suggest different scaling considerations in order to minimise or avoid the overflow situations in the calculation process. We apply the scaling process at the input of the iterative section to $w_1$ and $w_2$, and not to the input signal, because this has a negative impact on the SNR, when they are reintroduced into the Goertzel algorithm mapping module (see figure \ref{fig:gf_mapping}) by an arithmetic shift (saving FPGA resources), and is configurable by software. The scaling applied is:

\begin{equation}
    ASF = \frac{4N}{\pi}
\end{equation}

where $ASF$ stands for \textit{Arithmetic Scaling Factor} and $N$ is the size of the window. This result is then converted to a defined number of bits to be shifted doing a ceiling of the following expression: 

\begin{equation}
    ASF_{bits} = \lceil \log_{2}(ASF) \rceil
\end{equation}

\subsection{Signal-to-Noise Ratio (SNR)}

The selected ADC samples at \SI{1}{GSPS} and has an Effective-Number-of-Bits (ENOB) of \SI{10.3}{\bit} at $f_{in}$ = \SI{450}{\mega\hertz}, which is used as the starting data width for the following calculations, and a protected aliasing bandwidth of \SI{192.5}{\mega\hertz} for $R_{ADC} = 4$ ($R_{ADC}$ is the decimation ratio configured in the ADC). The estimated SNR is calculated as:

\begin{equation}
	SNR = 6.02N + 1.76 dB + 10\log_{10}(\frac{Fs}{2BW})
	\label{eq:snr_adc}
\end{equation}

The decimation stage adds a Processing Gain (PG) to the base SNR of the ADC of $\approx$ \SI{4.145}{\dB} (in accordance with the datasheet specification of 4 dB), giving an SNR $\approx$ \SI{67.91}{\dB} which requires 11 Bits for data width according to eq. \eqref{eq:snr_adc}. Taking into account the impact of the clock jitter on the SNR, we need to correct the previous value using the following expressions:

\begin{eqnarray}
	SNR_{jitter} = -20\log_{10}(2\pi f_{in} t_{j}) \label{eq:snr_adc_a}\\
	SNR_{0} = 10\log_{10}[10^{(-\frac{SNR_{ADC}}{10})} + 10^{(-\frac{SNR_{jitter}}{10})}] \label{eq:snr_adc_b}
\end{eqnarray}

This gives an initial SNR\textsubscript{0} $\approx$ \SI{67.60}{\dB} which also requires 11 Bits for $t_j = 55$ fs rms (for AD9680). Afterwards, the second decimation stage takes place and introduces a new PG relative to the effective BW of this stage, which is about \SI{29}{\mega\hertz}. The PG is defined by the term: $10\log_{10}(\frac{Fs}{2BW})$ in eq. \ref{eq:snr_adc}. After the second decimation stage, we have a SNR\textsubscript{1} $\approx$ \SI{74.01}{\dB} which requires 12 bits as data width. Finally, the GF is applied while the PG depends on the size and type of the selected window: 

\begin{table}[H]
	\caption{Window functions PG, final SNR  and final number of bits, for window size of $N = 256$ samples.}
	\label{tab:window_list_pg}
	\begin{center}
		\begin{tabular}{cccc}
			\toprule 
			Window Function & PG [dB] & SNR [dB] & NBits\\ 
			\midrule 
			Modified Flat-top & 15.34 & 89.36 & 15\\ 
			Dolph-Chebyshev & 18.42 & 92.43 & 16\\ 
			Rectangular &  21.55 & 95.56 & 16\\ 
			\bottomrule  
		\end{tabular} 
	\end{center}
\end{table}

\begin{figure}[H]
	\centering
	\includegraphics[width=0.9\columnwidth]{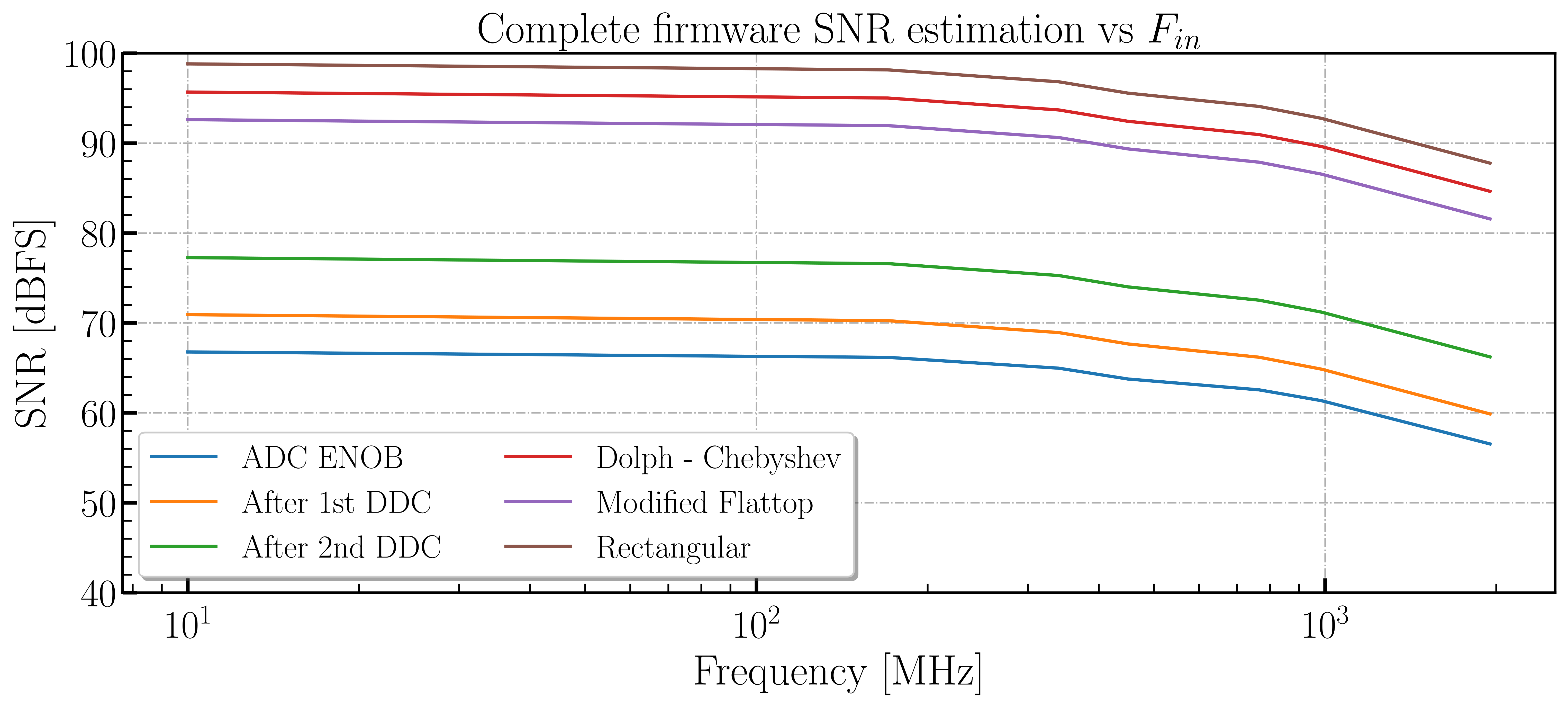}
	\caption{Signal-to-Noise Ratio as a function of the input frequency. ADC ENOB refers to the SNR of the ADC before the first decimation stage, considering the jitter contribution. The following curves for the DDCs are obtained by adding the Processing Gain. Finally, three different windows are evaluated.}
	\label{fig:snr_performance}
\end{figure}

From this analysis, we can estimate the SNR performance of the entire channelizer as a function of the input frequency (see figure \ref{fig:snr_performance}) and the necessary number of bits for each stage in order to preserve the expected SNR (see table \ref{tab:snr_list}).

\begin{table}[H]
	\begin{center}
		\caption{Data width after each stage. \textit{ADC Converter} refers to the ADC block within the chip previous to the internal DDC. \textit{ADC DDC} refers to the internal DDC of the ADC. \textit{DDC firmware} refers to the described hardware within the FPGA.}
		\label{tab:snr_list}
		\begin{tabular}{cccc}
			\toprule 
			ADC Converter & ADC DDC & DDC (firmware) & Goertzel Filter\\ 
			\midrule 
			16 & 16 & 18 & 32\\ 
			\bottomrule  
		\end{tabular} 
	\end{center}
\end{table}

%% file: sections/validation.tex
\section{Validation}
\label{sec:validation}

\subsection{Channelizer Frequency Response}

We performed the profiling of the designed DDC and GF by means of frequency sweeps. For these measurements, we looped inside the FPGA the signal generation module output directly to the channelizer, in order to have a clean characterisation of the channelizer without the converters contribution. 

\begin{figure}[H]
	\centering
	\includegraphics[width=1\columnwidth]{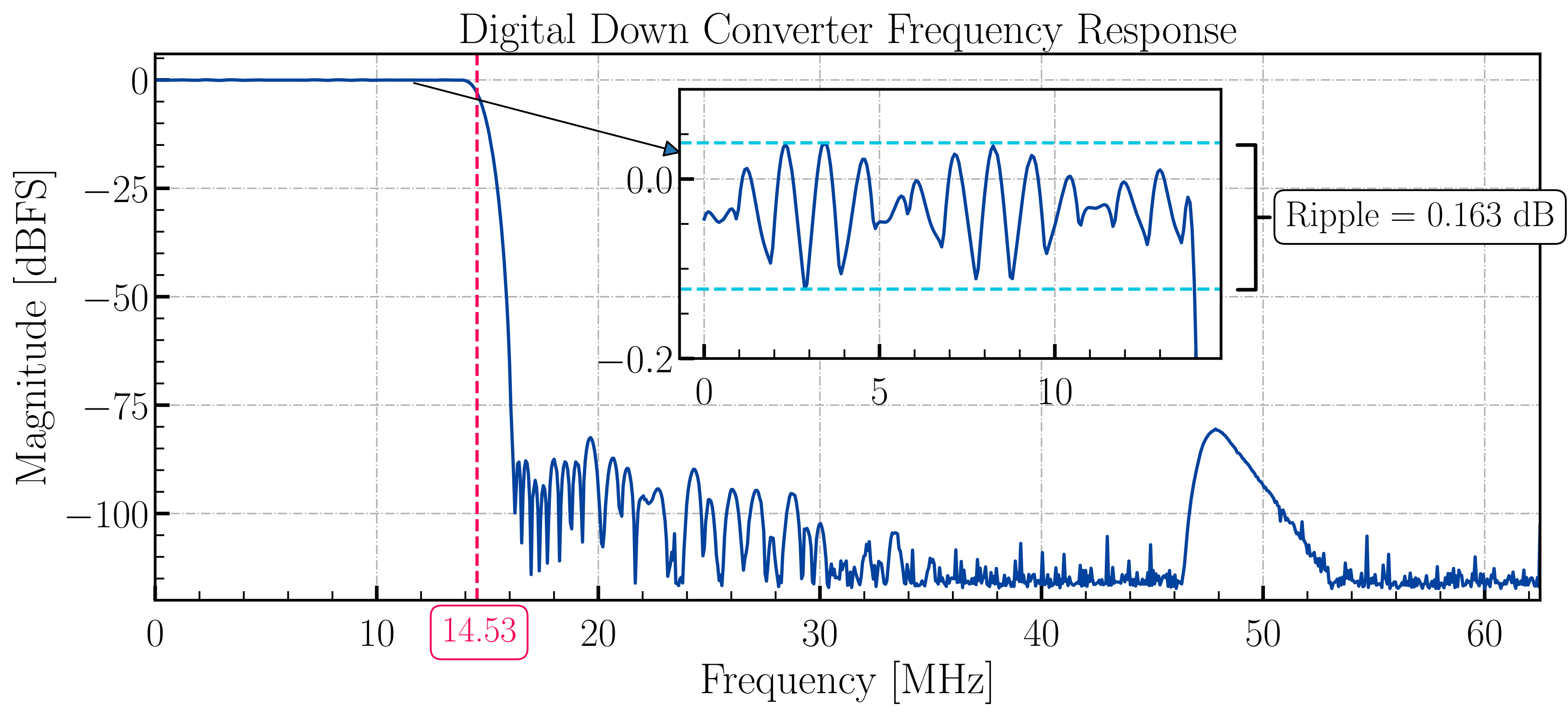}
	\caption{Digital Down Converter frequency response. A maximum ripple of \SI{0.163}{\dB} in the band pass and a \SI{3}{\dB} cut-off frequency at \SI{14.53}{\mega\hertz} are observed.} 
	\label{fig:ddc_response_validation}
\end{figure}

For the DDC, the frequency sweep was from \SI{15.625}{MHz} to \SI{78.125}{MHz}, with a $\Delta f$ of \SI{65}{kHz}, at \SI{0}{dBFS}. Figure \ref{fig:ddc_response_validation} shows the results which are consistent with the DDC simulations and expected magnitude response shown in figure \ref{fig:ddc_chain_model}. The discrepancy between the simulation and the implementation, is mainly related with the quantization noise: while the simulation was performed in Python using full precision data, floating point arithmetic, the implementation in the FPGA uses only 16 bits with a fixed point arithmetic. A second contribution is related to the spectral resolution.

\begin{figure}[H]
	\centering
	\includegraphics[width=1\columnwidth]{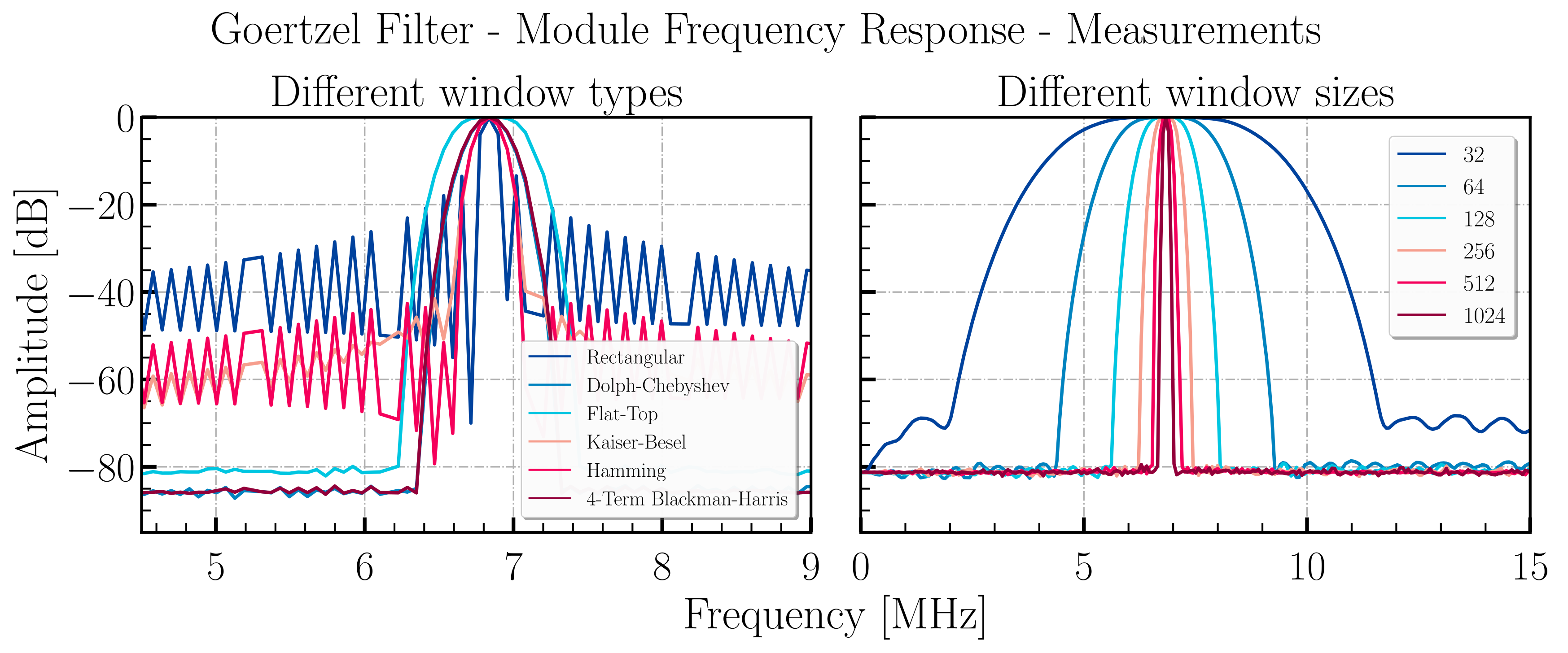}
	\caption{\textbf{Left:} channelizer frequency response for different window types for a 256 samples window size. \textbf{Right:} different window sizes for a Flat-Top window type.}	
	\label{fig:dsp_channelizer_response}
\end{figure}

For the GF we carried out a frequency sweep from \SI{0}{Hz} to \SI{15}{MHz}, with a $\Delta f$ of \SI{65}{kHz}, at \SI{0}{dBFS}. All the implemented filters will differ in their centre frequency (the DFT bin to detect), but will have the same profile. This measurement is shown in figure \ref{fig:dsp_channelizer_response}, which shows a consistent profile with the window function used. 

zz

\subsection{Demodulation capabilities}

\begin{figure}[H]
	\centering
	\begin{minipage}[c]{0.4\textwidth}
            \centering
		\includegraphics[width=0.85\columnwidth]{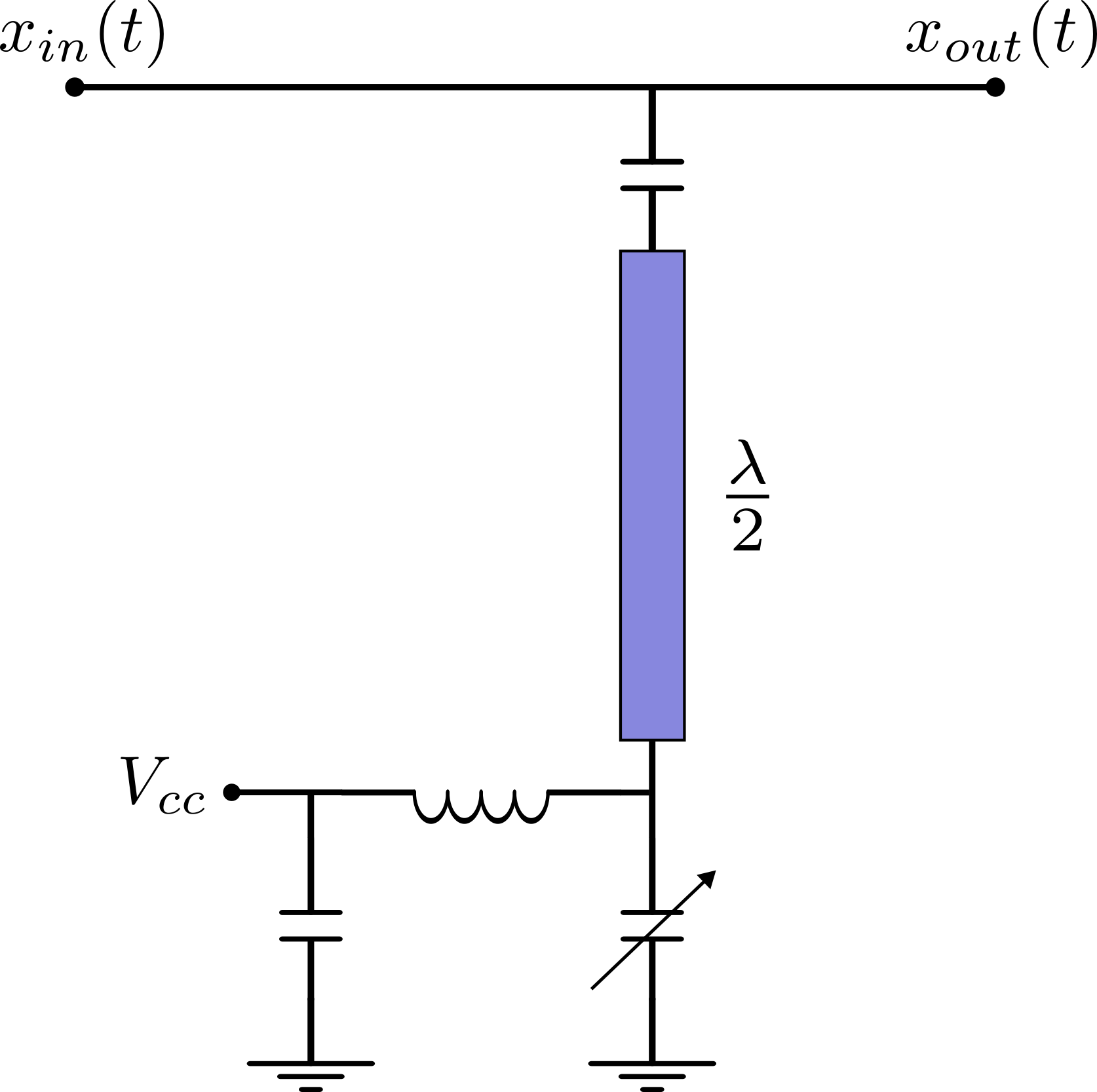}
	\end{minipage}\hfill
	\begin{minipage}[c]{0.55\textwidth}
		\includegraphics[width=1.1\columnwidth]{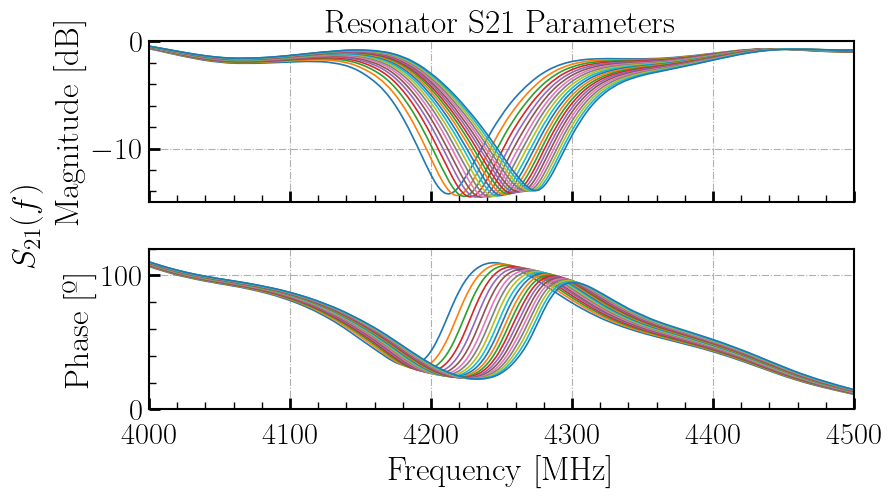}
	\end{minipage}
    \caption{\textbf{Left}: resonator's schematic. The right and left terminals are the input/output (bi-directional), and $V_{cc}$ terminal is the Voltage Control which defines the resonance frequency of the device. The voltage range for this terminal goes from \SI{0}{Vdc} to \SI{10}{Vdc}. \textbf{Right}: resonator's S21 parameter, characterized using a Vector Network Analyzer for different Vdc values, from \SI{0}{Vdc} to \SI{10}{Vdc} with a step of \SI{0.5}{Vdc}.}
	\label{fig:the_resonator}
\end{figure}

To test the demodulation capabilities, we implemented a resonator at room temperature using a varicap, which allows us to modify the resonance frequency of the device by varying the voltage applied to it, see figure \ref{fig:the_resonator}. Doing the calculation of a single bin of the DFT, is similar to an AM Demodulation process due to the multiplication with the $e^{-j2\pi k}$ factor, as was presented in section \ref{sec:goertzeltheory}. 

As can be seen in figure \ref{fig:the_resonator} the resonator operates between \SI{4.2}{GHz} and \SI{4.3}{GHz}, and as the digital backend works in BB in the range of \SI{-500}{MHz} to \SI{500}{MHz} (complex signals, IQ), a \textit{Radio Frequency Frontend} (RF-FE) was required to perform the up-conversion and down-conversion process for the generated BB spectrum, as was depicted in figure \ref{fig:read_out_system}.

For the tests in subsections \ref{sec:mod_type_i} and \ref{sec:mod_type_ii}, the Digital Backend generated a tone at \SI{-239}{\mega\hertz} with a power of \SI{-10}{\dBm} in order to monitor the behaviour of the resonators.

\subsubsection{Modulation Type I: Amplitude Modulation}
\label{sec:mod_type_i}

\begin{figure}[H]
	\centering
        \includegraphics[width=1\columnwidth]{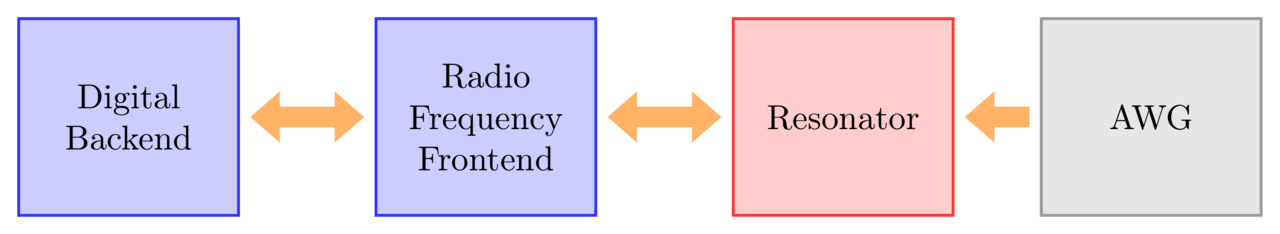}
        \caption{Experimental setup for AM modulation: placing a modulating signal (sinusoidal, triangular, square, etc.) in the resonator's voltage control terminal using an Arbitrary Wave Generator (AWG) makes its resonance frequency to change. To reach a full dynamic range, the used signals have a DC offset of \SI{5}{Vdc}. 
        However, for these first tests we worked within the range of \SI{5}{Vdc} $\pm$ \SI{2}{Vdc}.}
        \label{fig:measurement_process_1}
\end{figure}

The main goal of this first experiment is to recover the amplitude modulated signal, which corresponds to an AM demodulation process. The experimental setup is sketched in figure \ref{fig:measurement_process_1}. 

\begin{figure}[H]
	\centering
	\includegraphics[width=\textwidth]{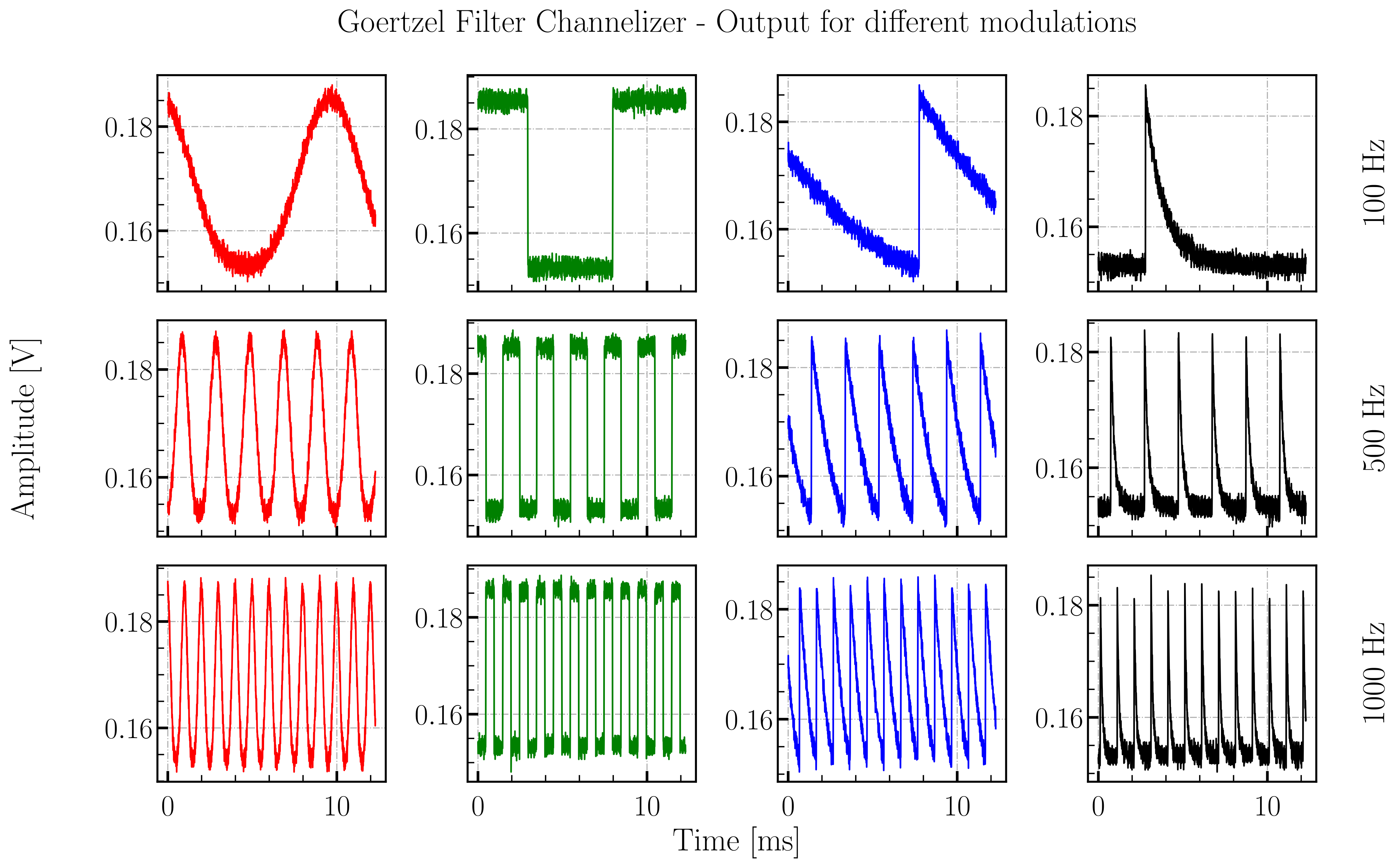}
	\caption{AM Demodulation from the resonator: GF output for different cases.}		
	\label{fig:dsp_am_modulation}
\end{figure}

\begin{figure}[H]
	\centering
	\includegraphics[width=\textwidth]{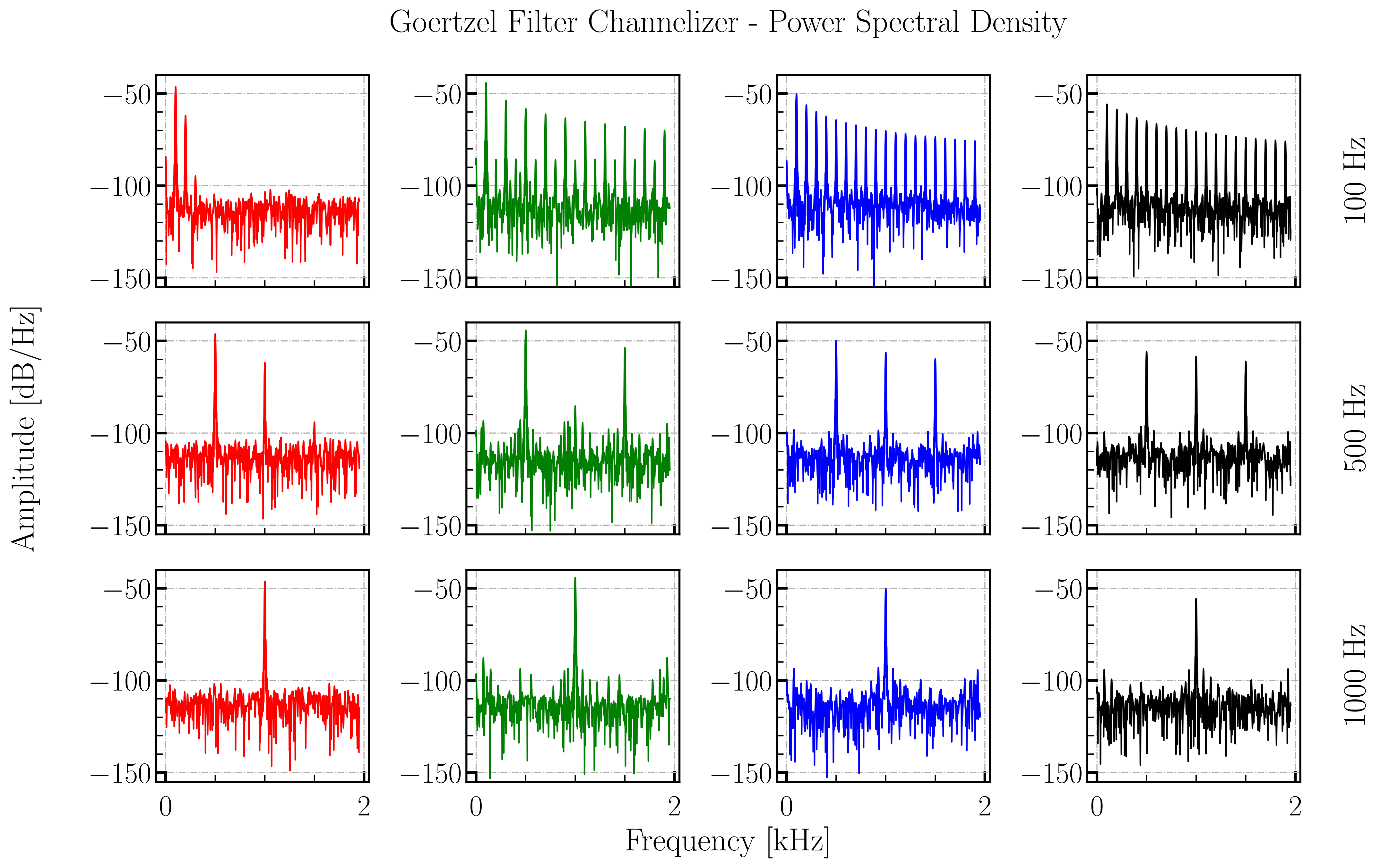}
	\caption{AM Demodulation from the resonator: Power Spectral Density (PSD) of figure \ref{fig:dsp_am_modulation} signals.}		
	\label{fig:dsp_am_modulation2}
\end{figure}

We generated different signals (sinusoidal, square, ramp and pulse), at different frequencies (from \SI{100}{Hz} to \SI{50}{kHz}) in order to validate the Goertzel Filter's bandwidth and performance. Some of them are shown in figures \ref{fig:dsp_am_modulation} and \ref{fig:dsp_am_modulation2}.

\subsubsection{Modulation Type II: Amplitude + Phase Modulation}
\label{sec:mod_type_ii}

As was explained in Section \ref{sec:readconcept}, working with \uMUX~demands the use of the Flux-Ramp Modulation technique \cite{Mates2012}. This means that the problem can be analysed as one where both AM and PM modulation are present in the carrier signal. We generated a PM modulated signal stimulating our resonator. This scenario represents the AM   , and  We generated the same signals stimulating our resonator as before, but added a PM modulation to this signal. The aim was to retrieve this PM modulated signal, which is the one carrying the information from the sensors. The demodulation method is based on a typical I/Q demodulation technique and performed afterwards in a python script:

\begin{equation}
    x'(t) = x(t) * w(t)
\end{equation}

\begin{equation}
	\phi = \arctan \left[ \frac{\sum x'(t)\cos(w_c t)}{\sum x'(t)\sin(w_c t)} \right]
\end{equation}

where $x(t)$ is the output of the GF and $w(t)$ is a window function which performs a filtering using a Dolph-Chebyshev configured to attenuate the Highest Side Lobe \SI{-200}{dB}). The use of a Hilbert transform as an alternative to this method is also feasible. For the experimental setup, a \SI{30}{kHz} sinusoidal signal emulating the SQUID response was generated; and on top of this, a \textit{phase modulation} of \SI{60}{º} at \SI{200}{Hz} of 3 different types (sinusoidal, triangular and square) was implemented. The following figures show the results:

\begin{figure}[H]
	\centering
	\includegraphics[width=0.9\textwidth]{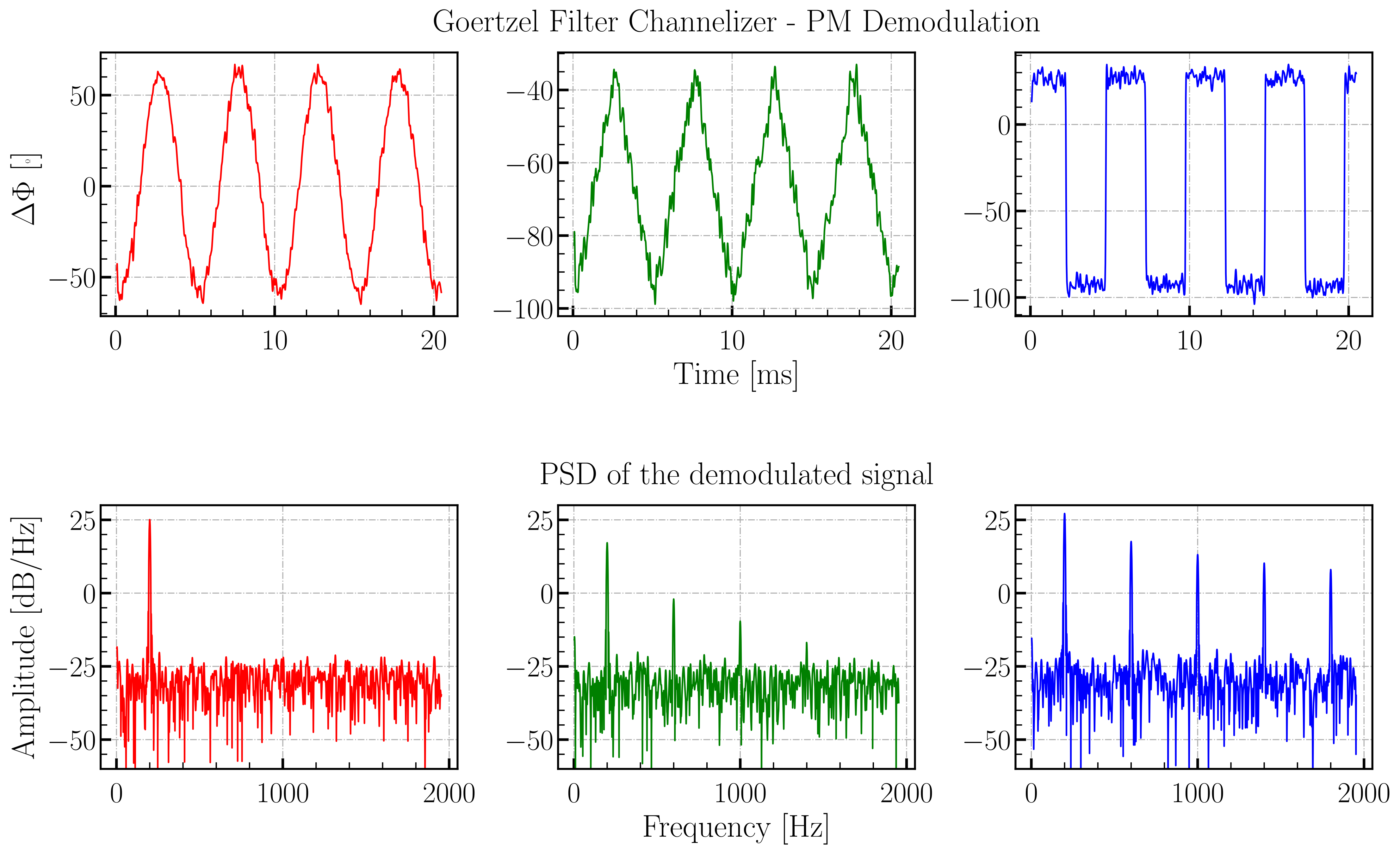}
	\caption{PM demodulation from the demodulated AM signal, emulating a detector signal. \textbf{Top:} total phase shift of \SI{120}{º} as expected. \textbf{Bottom:} Power Spectral Density of the top plot, showing the consistency of the visualized signal and its spectrum; where the main component is located at \SI{200}{Hz}.}		
	\label{fig:dsp_am_pm_modulation}
\end{figure}

%% file: sections/conclusions.tex
\section{Summary}
In this paper we have demonstrated that by estimating a bin of the DFT using a Goertzel Filter, it is possible to retrieve the desired scientific data from an FDM readout system applied to \uMUX~ using the FRM technique. This work presents an in-depth study of the GF mathematics and a simulation framework of the entire channellizer, showing an excellent agreement between the simulated and experimental results. A resource-efficient, feasible and reliable mapping to a state-of-the-art FPGA is presented, taking into account many effects of this process such as arithmetic scaling (bit growth control in DDC and GF), fixed-point arithmetic and window function amplitude compensation. 
The final channelization step, performed with the GF, shows the expected dependence on the window function used to isolate adjacent channels with more than \SI{80}{\dB}, which is sufficiently good for the requirements of our experiments. An interesting feature of this channelizer is the coarse and fine tuning to detect the required signals, thanks to the combination of DDC and GF. Furthermore, it is possible to double the density of detectors and use the same firmware implementation by simply rearranging the DDC configuration on the fly, even though this represents the loss of the half of the bandwidth. Finally, the throughput achieved in this implementation is related to the configured window size, N, by \SI{31.25}{MSPS} / N. In our measurements we used a window size of 256 samples and obtained a throughput of \SI{122.07}{KSPS}.